\documentclass[11pt]{article}

\usepackage[top = 30 mm, bottom = 30mm, left = 20mm, right=20mm]{geometry}
\usepackage{amsmath, amssymb}
\usepackage[english]{babel}
\usepackage[utf8]{inputenc}
\usepackage{latexsym}
\usepackage{graphicx}
\usepackage{lmodern}
\usepackage{jheppub}
\usepackage{slashed}
\usepackage{charter}
\usepackage{xcolor}
\usepackage{color}
\usepackage{float}
\usepackage{bm}
\usepackage{braket}
\usepackage{dsfont}
\usepackage{cancel}

\numberwithin{equation}{section}

\newcommand{\be}{\begin{equation}}
	\newcommand{\ee}{\end{equation}}
\newcommand{\bear}{\begin{array}}
	\newcommand{\eear}{\end{array}}

\title{\Large Cosmological perturbations from five-dimensional inflation}

\author[a,b,c]{Ignatios Antoniadis}

\author[c]{Jules Cunat}
\author[c]{Anthony Guillen}

\affiliation[a]{Simons Center for Geometry and Physics\\
Stony Brook University, Stony Brook, NY 11794, USA}

\affiliation[b]{Center for Cosmology and Particle Physics, Department of Physics\\New York University, 726 Broadway, New York, NY 10003, USA}

\affiliation[c]{Laboratoire de Physique Th\'eorique et Hautes \'Energies - LPTHE\\Sorbonne Universit\'e, CNRS, 4 Place Jussieu, 75005 Paris, France}
 
\emailAdd{antoniad@lpthe.jussieu.fr}
\emailAdd{jcunat@lpthe.jussieu.fr}
\emailAdd{aguillen@lpthe.jussieu.fr}

\date{} 

\abstract{It was recently proposed that five-dimensional inflation can relate the causal size of the observable universe to the present weakness of gravitational interactions by blowing up an extra compact dimension from the microscopic fundamental length of gravity to a large size in the micron range, as required in the Dark Dimension proposal. Here, we compute the power spectrum of all primordial fluctuations emerging from a 5-dimensional inflaton in a slow-roll region of its potential, showing an interesting change of behaviour at large scales corresponding to angles larger than about $10$ degrees in the sky.}

\begin{document}

\maketitle

\section{Introduction}

In a recent work~\cite{Anchordoqui:2023etp}, it was proposed that compact extra dimensions can acquire large size by higher dimensional inflation, thus connecting two large hierarchies in particle physics and cosmology: the weakness of gravitational interactions when compared to the Standard Model gauge interactions, and the largeness of the observable universe relative to our past causal horizon. Indeed, large extra dimensions compared to the fundamental length of gravity~\cite{Arkani-Hamed:1998jmv, Antoniadis:1998ig}, such as the string or species scale~\cite{Dvali:2007hz}, can account for the observed weakness of gravitational force on our 3-brane universe, localised in the extra dimensions. On the other hand, their size may have become large due to uniform higher dimensional inflation that expanded also the size of our observable world~\cite{Anchordoqui:2022svl}. Note that the underlying physical mechanism for producing the two hierarchies has the same origin: large extra dimensions decrease the strength of gravity because the gravitational field can spread into the compact space, while a higher dimensional universe needs less number of e-folds of expansion to solve the horizon problem.\\

This proposal becomes particularly attractive in the case of one extra dimension of micron size corresponding to a five-dimensional (5D) gravity scale of order $10^9$ GeV. The reason is that only in this case higher dimensional inflation can also lead to an approximate scale invariant power spectrum of primordial density perturbations~\cite{Harrison:1969fb, Zeldovich:1972zz} consistent with observations of the cosmic microwave background (CMB) anisotropies~\cite{Anchordoqui:2023etp}. Indeed, the size of extra dimensions introduces a change of behaviour of 2-point cosmological correlators at a critical distance around the compactification scale, implying violations of scale invariance in the power spectrum at larger wave lengths, while at shorter distances scale invariance is maintained on our 3-brane universe upon summation over the Kaluza-Klein (KK) modes. \\

On the other hand, experimental precision measurements of the power spectrum are valid for angles less than about $10$ degrees, corresponding to multiple moments $l\gtrsim 30$, or equivalently to distances less than Mpc at the CMB, or Gpc today~\cite{Planck:2018jri}. Extrapolating this distance back in the past, using radiation dominated expansion and changing units to higher dimensional Einstein frame, one finds that the Mpc becomes a micron.  This selects the case of one extra dimension of micron size as the only possibility where higher dimensional inflation can also generate nearly scale invariant spectrum of primordial perturbations consistent with observations. It can also be implemented in the Dark Dimension proposal for the cosmological constant~\cite{Montero:2022prj} based on the distance/duality conjecture of the Swampland program~\cite{Vafa:2005ui, Ooguri:2006in}.\\

In this work, we compute the power spectrum of cosmological fluctuations in the slow-roll approximation of a five-dimensional (5D) inflaton, generalising the standard 4D formulae. In terms of physical degrees of freedom, besides the 5D inflaton, one has the 5D graviton that has five polarisations. From a 4D perspective, these lead to six 4D zero-modes consisting of two scalars (the inflaton and the radion), the spin-2 graviton with 2 polarisations (B-modes), and a vector, as well as a tower of massive spin-2 KK excitations. In a standard brane-world construction, the fifth dimension is compactified on a line interval $S^1/\mathbb{Z}_2$ and the vector 0-mode is projected away by the $\mathbb{Z}_2$ action. Moreover, the Standard Model is localised on a 3-brane located at the origin of the extra dimension. \\

Cosmological perturbations observed in our worldbrane are characterised by 3D wave lengths entering the horizon after the end of inflation. Evaluating them at the position of the brane amounts to performing a summation over all internal KK momenta. As mentioned above, this summation is crucial for obtaining nearly scale invariant power spectra at `small' distance scales. Moreover, density perturbations receive contributions from two scalars: the 5D inflaton and the radion together with the KK excitations of the scalar polarisation of the 5D graviton. However, as the primordial gravity waves in four dimensions, the contribution of the later is suppressed by the slow-roll parameter. The same parameter suppresses also the vector fluctuations, as well as the isocurvature entropy perturbations, generated as in multifield 4D inflation models. Here however, these perturbations appear at second order. \\

A general property of all primordial fluctuations from 5D inflation is a change of behaviour at large angles leading to more power compared to the 4D case, with a nearly vanishing spectral index. This prediction can in principle be confronted in future CMB observations with improved precision at low multipole moments. \\

The outline of our paper is the following. In Section 2, we recall for pedagogical reasons the computation of scalar and tensor perturbations within single field 4D inflation, in a way that can be generalised in five dimensions, as we do in the following sections: scalar perturbations in Section 3, isocurvature, entropy perturbations in Section 4 and tensor and vector perturbations in Section 5. Finally, Section 6 contains our conclusions. In all of this paper, we use natural units and the mostly-plus convention for the signature of the metric.

\section{Perturbations in four dimensions}

Before extending it to five dimensions, we review the standard computation of the scalar and tensor perturbations in four dimensions \cite{Riotto:2002yw, Baumann:2009ds, Langlois:2010xc}. It starts from perturbing the metric $g_{\mu\nu}$ and the inflaton $\phi$ around a time-dependent background, solution of the equations of motion in an approximate flat (slow-roll) region of the scalar potential:
\begin{equation}
    \phi(t, x) = \bar\phi(t) + \delta\phi(t,x),
\end{equation}
and
\begin{equation}\label{eq:pre_7}
ds^2 = -(1+2\Phi)dt^2 + 2a(t)B_idx^idt + a^2(t)((1 - 2\Psi)\delta_{ij} + E_{ij})dx^idx^j.
\end{equation}
We can further decompose $B_i$ and $E_{ij}$ into scalar, vector and tensor modes
\begin{equation}
    B_i = \partial_i B + C_i
    \qquad\text{and}\qquad
    E_{ij} = 2\partial_{i}\partial_j E + 2\partial_{(i}E_{j)} + h_{ij}.
\end{equation}
So in the end, we have five scalar perturbations $(\delta\phi, \Phi, \Psi, B, E)$; two vector perturbations $(C_i, E_i)$ and one tensor perturbation $h_{ij}$. They transform under diffeomorphisms
\begin{equation}
    t \rightarrow t + \xi^t
    \quad\text{and}\quad
    x^i \rightarrow x^i + \delta^{ij}\partial_j\xi^x,
\end{equation}
as
\begin{equation}\label{eq:pre_2}
    \delta\phi \rightarrow \delta\phi - \dot{\phi}\xi^t,\quad
    \Phi \rightarrow \Phi - \dot\xi^t,\quad
    \Psi \rightarrow \Psi + H\xi^t,\quad
    B \rightarrow B + a^{-1}\xi^t - a\dot\xi^x,\quad
    E \rightarrow E - \xi^x.
\end{equation}
for the scalar perturbations and similarly for the vectors and tensors.
Inflaton perturbations also induce perturbations of the energy-momentum tensor $T^\mu_\nu$, parameterised as
\begin{equation}
    \delta T^0_0 = -\delta\rho, \qquad \delta T^0_i = a \partial_i\delta q, 
    \qquad T^i_j = \delta^i_j\delta p + \Sigma^i_j,
\end{equation}
where the energy-momentum tensor of the scalar field is given by
\begin{equation}
    T^\mu_\nu = g^{\mu\rho}\partial_\nu\phi\partial_\rho\phi - \delta^\mu_\nu(1/2(\partial\phi)^2 + V).
\end{equation}
In particular, the anisotropic stress tensor $\Sigma^i_j$ vanishes for a single field. The other perturbations of the energy-momentum tensor $(\delta\rho, \delta q,  \delta p)$ also transform under \eqref{eq:pre_2}, and they can be combined with perturbations of the metric into gauge invariant quantities, like the comoving curvature perturbation
\begin{equation}
    \mathcal{R} = \Psi - \frac{H}{\bar\rho + \bar p}\delta q,
\end{equation}
which is the kind of quantity that we are interested in. For completeness, we recall the Friedmann equations governing the background evolution
\begin{equation}
    3H^2 = 1/2 \dot\phi^2 + V \quad \text{and} \quad
    \dot H = -1/2\dot\phi^2.
\end{equation}

\subsection{Scalar perturbations}

Using a transformation with $\xi^t$ in  \eqref{eq:pre_2}, we can put ourselves in the gauge $\delta\phi = 0$. In this gauge, $\delta q = 0$ as well and $\mathcal{R} = \Psi$. We can also choose $E = 0$ since it transforms with $\xi^x$. Even if $\delta\phi = 0$, the energy and pressure density perturbations $(\delta\rho, \delta p)$ are not vanishing at first order
\begin{equation}
\delta\rho = -1/2g^{00}\dot\phi^2 \simeq -\dot\phi^2\Phi \qquad\text{and}\qquad\delta p \simeq -\dot\phi^2\Phi.
\end{equation}
The other components of the energy momentum tensor, however, remain unperturbed.\\

From there, the procedure is quite simple. First, we write the Einstein equations for the metric \eqref{eq:pre_7} in the gauge $\delta\phi = E = 0$; in Fourier space, they take the following form
\begin{eqnarray}\label{eq:pre_19}
    3H(-\mathcal{\dot R} + H\Phi) - k^2/a^2(\mathcal{R} + a H B) &=& {1\over 2}\dot\phi^2\Phi\nonumber\\
    -\mathcal{\dot R} + H\Phi &=& 0\nonumber\\
    -\mathcal{\ddot R} - 3H\mathcal{\dot R} + H\dot\Phi + (3H^2 + 2\dot H)\Phi &=& -{1\over 2} \dot\phi^2\Phi\nonumber\\
    (\partial_t + 3H)B/a + (\mathcal{R} + \Phi)/a^2 &=& 0
\end{eqnarray}
the first one is the $00$ component, the second is the $0i$, and the last two come from the $ij$ components. Now, the first two equations can be combined to obtain
\begin{equation}
    \Phi = \frac{\mathcal{\dot R}}{H}
    \quad\text{and}\quad
    B = -\frac{2k^2H\mathcal{R} + a^2\mathcal{\dot R}\dot\phi^2}{2ak^2H^2},
\end{equation}
so we see that $\Phi$ and $B$ are not dynamical fields but solutions of constraint equations. Plugging their expressions in the two remaining equations of \eqref{eq:pre_19} yields
\begin{equation}\label{eq:pre_8}
    \mathcal{\dot R}(2\dot H + \dot\phi^2) = 0,
\end{equation}
and 
\begin{equation}\label{eq:pre_9}
    a^3H\dot\phi^2\mathcal{\ddot R} + a^3\dot\phi(3H^2\dot\phi - 2\dot H\dot\phi + 2H\ddot\phi)\mathcal{\dot R} - 2k^2H(2\dot a H - 2aH^2 + a \dot H)\mathcal{R} = 0.
\end{equation}
Thus, \eqref{eq:pre_8} reduces to an equation on the background which is trivially satisfied. The other one, \eqref{eq:pre_9}, can be rewritten as
\begin{equation}\label{eq:pre_24}
    \frac{a^3\dot\phi^2}{H^2}\mathcal{\ddot R} + \frac{d}{dt}\left(\frac{a^3\dot\phi^2}{H^2}\right)\mathcal{\dot R} + \frac{a\dot\phi^2}{H^2}k^2\mathcal{R} = 0,
\end{equation}
which can be identified as  equation of motion from the action
\begin{equation}\label{eq:pre_241}
    \mathcal{S} = \frac{1}{2}\int d^4x\frac{a^3\dot\phi^2}{H^2}((\mathcal{\dot R})^2 - a^{-2}(\partial_i\mathcal{R})^2).
\end{equation}
This action takes a better form defining  $z = a\dot\phi H^{-1}$ and $v = z\mathcal{R}$, it reads in conformal time
\begin{equation}\label{eq:pre_3}
    \mathcal{S} = \frac{1}{2}\int d\tau d^3x\left((v')^2 + (\partial_iv)^2 + \frac{z''}{z}v^2\right)
\end{equation}
In Fourier space, the action \eqref{eq:pre_3} leads to the Mukhanov-Chibisov equation~\cite{Mukhanov:1981xt}
\begin{equation}\label{eq:pre_4}
    v''_k + \left(k^2 - \frac{z''}{z}\right)v_k = 0,
\end{equation}
which can be solved numerically or analytically once the background evolution is specified. \\

Around a slow-roll region of the scalar potential, we introduce the slow-roll parameters, along with the conformal Hubble rate $\mathcal{H} = a'/a$
\begin{equation}\label{eq:pre_290}
    \varepsilon = -\frac{\dot H}{H^2} = 1 - \frac{\mathcal{H}'}{\mathcal{H}^2}
    \qquad\text{and}\qquad
    \delta = -\frac{\ddot\phi}{H\dot\phi} = 1 - \frac{\phi''}{\mathcal{H}\phi'}.
\end{equation}
We can also introduce the second Hubble slow-roll parameter $\varepsilon_2 = \dot\varepsilon/(\varepsilon H) = -2\delta + 2\varepsilon$ as in \cite{Motohashi:2014ppa}, but $\delta$ is more convenient here. We also introduce the potential slow-roll parameters
\begin{equation}
    \varepsilon_V = \frac{1}{2}\biggl(\frac{V_{\phi}}{V}\biggr)^2 \quad\text{and}\quad \eta_V = \frac{V_{\phi\phi}}{V},
\end{equation}
where the $\phi$ index denotes a derivative with respects to $\phi$. In the slow-roll regime, it is easy to show that $\varepsilon \simeq \varepsilon_V$ and $\delta \simeq \eta_V - \varepsilon_V$. From the definition of $\varepsilon$, we can also get $\mathcal{H} \simeq -(1+\varepsilon)/\tau$. 
Noticing that $z = a\phi'\mathcal{H}^{-1}$, it is rather easy to obtain that $z'/z = \mathcal{H}(1+\varepsilon+\delta)$, and then
\begin{equation}\label{eq:pre_29}
    \frac{z''}{z} \simeq \mathcal{H}^2(2+2\varepsilon-3\delta) \simeq \frac{2+6\varepsilon-3\delta}{\tau^2},
\end{equation}
at first order in the slow-roll parameters (which amounts to assume that they are constants in time, i.e. their variations are of second order). Note that in the de Sitter limit $z''/z \simeq 2/\tau^2$, the general solution of equation \eqref{eq:pre_4} is
\begin{equation}
    v_k = c_1\biggl(1 - \frac{i}{k\tau}\biggr)\frac{e^{-ik\tau}}{\sqrt{2k}} + c_2\biggl(1 + \frac{i}{k\tau}\biggr)\frac{e^{ik\tau}}{\sqrt{2k}}.
\end{equation}
Then, Bunch-Davies boundary condition fixes $c_1 = 1, c_2 = 0$, and we get a scale invariant power spectrum for $\mathcal{R}$
\begin{equation}
    \mathcal{P}_\mathcal{R} = \frac{k^3}{2\pi^2}\frac{H^2}{a^2\dot\phi^2}|v_k|^2
     = \frac{1}{2\varepsilon}\left(\frac{H}{2\pi}\right)^2,   
\end{equation}
where we used $a(\tau) = -1/(H\tau)$, and the super horizon limit $k\tau \ll 1$.\\ 

If we don't neglect the slow-roll parameters, the general solution of \eqref{eq:pre_4} is given in terms of Bessel functions
\begin{equation}
    v_k = c_1\tau^{1/2}J_\nu(k\tau) + c_2\tau^{1/2}Y_\nu(k\tau)\,,
\end{equation}
where the order of the Bessel functions is given by $\nu=\sqrt{ 9/4 + 6\varepsilon - 3\delta}\simeq 3/2 + 2\varepsilon - \delta$. The constants $c_i$ are fixed again by the Bunch-Davies boundary condition
\begin{equation} \label{eq:pre_30}
    \lim_{\tau\rightarrow-\infty}v_k = \frac{\exp(-ik\tau)}{\sqrt{2k}},
\end{equation}
giving 
\begin{equation}   \label{eq:pre_301} 
    c_1 = \frac{1-i}{2}\sqrt{\frac{\pi}{2}}\mathrm{exp}\biggl(-\frac{1}{2}i\pi\nu\biggr) 
    \quad\text{and}\quad
    c_2 = -\frac{1+i}{2}\sqrt{\frac{\pi}{2}}\mathrm{exp}\biggl(-\frac{1}{2}i\pi\nu\biggr)\,.
\end{equation}
In the super horizon limit $\tau\to 0$, one obtains
\begin{equation}\label{eq:pre_27}
    \mathcal{P}_\mathcal{R} = \frac{1}{2\varepsilon}\biggl(\frac{H}{2\pi}\biggr)^2\biggl(\frac{k}{aH}\biggr)^{3-2\nu},
\end{equation}
which is almost the same result as before, up to the spectral tilt $n_\mathcal{R} - 1 = 2\delta - 4\varepsilon = 2\eta_V - 6\varepsilon_V$.

\subsection{Tensor perturbations}

Computing the power spectrum of tensor perturbations is straightforward. Following a similar procedure as for obtaining \eqref{eq:pre_3}, the second order action for the tensor perturbation $h_{ij}$ is
\begin{equation}\label{eq:pre_500}
    \mathcal{S}_h = \frac{1}{8}\int d^4x a^2((h'_{ij})^2 - (\partial_l h_{ij})^2).
\end{equation}
Going in Fourier space, where the two polarisations of $h_{ij}$ are described by the modes $h_k^s$, and defining $2v_k^s = a h_k^s$, we obtain an equation of motion very similar to \eqref{eq:pre_4}
\begin{equation}\label{eq:pre_5}
    (v_k^s)'' + \left(k^2 - \frac{a''}{a}\right)v_k^s = 0.
\end{equation}
At first order in the slow-roll parameters, we have $a''/a \simeq (2+3\varepsilon)/\tau^2$, and we can thus repeat the previous analysis with $\nu = \sqrt{9/4 + 3\varepsilon}\simeq 3/2 + \varepsilon$, to obtain the result from the scalar spectrum using the new value of $\nu$ and omitting the pre-factor $H^2/\dot\phi^2$. The resulting tensor power spectrum is
\begin{equation}\label{eq:pre_38}
    \mathcal{P}_h = \frac{2H^2}{\pi^2}\biggl(\frac{k}{aH}\biggr)^{-2\varepsilon},
\end{equation}
where a factor $2$ comes from the fact that there are two polarisations, and an extra factor of $4$ coming from the normalisation in $2v_k^s = a h^s_k$. The most notable difference from the scalar spectrum $\mathcal{P}_\mathcal{R}$ is the absence of the factor of $\varepsilon$ that previously came from the $H^2/\dot\phi^2$ in the relation between $v_k$ and $\mathcal{R}$. This suppresses the ratio $r$ of the tensor-to-scalar perturbations by the slow-roll parameter, $r\simeq 16\varepsilon$.

\section{Perturbations in five dimensions}

We now consider the five-dimensional (5D) case that we are actually interested in. We start with reviewing the background evolution, before computing the perturbations as in the previous section.

\subsection{Background evolution}

The metric background we consider is nearly five-dimensional de Sitter with one compact space dimension. In this section, we will follow the notations of \cite{Anchordoqui:2023etp}. The hats are here to make the difference between 5D and 4D quantities, as this will be important in the discussion above Figure \ref{fig:1}. The 5D metric is therefore given by the line element
\begin{equation}\label{eq:pre_6}
    ds^2 = -d{\hat t}^2 + \hat{a}^2({\hat t})dx^2 + R^2({\hat t})dy^2,
\end{equation}
with periodicity $y \sim y + 2\pi$, so that $R({\hat t})$ is the radius of the compact circular dimension. At this stage, both $\hat{a}({\hat t})$ and $R({\hat t})$ are expanding exponentially in a uniform way, with $\hat{a}({\hat t}) = e^{H{\hat t}}$ and $R({\hat t}) = R_0e^{H{\hat t}}$. We can also define the conformal time $\tau = -H^{-1}\exp(-H{\hat t})$, for which $\hat{a}(\tau) = -1/(H\tau)$ and $R(\tau) = -R_0/(H\tau)$. With the conformal time, the metric \eqref{eq:pre_6} becomes
\begin{equation}
\label{conformallyflatmetric}
    ds^2 =\hat{a}^2(\tau)(-d\tau^2 + dx^2 + R_0^2dy^2).
\end{equation}

As mentioned in the introduction, the extra dimension is compactified on an interval $S^1/\mathbb{Z}_2$ with $\mathbb{Z}_2$ the parity symmetry $y\to -y$, leading at two fixed points at the ends, where branes with our observable universe can be localised.
In the string theory context, the branes are described by open strings on D-branes localised in the extra dimension with the matching conditions automatically satisfied due to the requirement of tadpole cancellation in the presence of (negative tension) orientifolds. The overall time dependence arises from the bulk cosmological constant and amounts of considering a slice of 5D de Sitter spacetime along the compact $y$-direction.\footnote{Of course, in the string theory context, this is approximate since an exact stable dS solution may not exist.} This should be contrasted to the more general warped cosmological solutions in the braneworld context studied in the past with less isometries~\cite{Lukas:1998qs, Binetruy:1999hy, Lukas:1999yn, Binetruy:1999ut}, or in an anti-de Sitter bulk~\cite{Randall:1999ee, Randall:1999vf}. \\

\newpage

During inflation, as in 4D, the background is not exactly de Sitter, but quasi de Sitter, sourced by a canonically normalised and minimally coupled scalar field, with the following equation of motion
\begin{equation}\label{eq:pre_21}
    \phi'' + 3\mathcal{H}\phi' + \hat{a}^2(\tau)\frac{dV}{d\phi} = 0,
\end{equation}
where from here onwards, we stick to conformal time and denote by $\mathcal{H}$ the conformal Hubble rate, $\mathcal{H} = \hat{a}'/
\hat{a}$. The 5D Friedmann equations read
\begin{equation}\label{eq:pre_39}
    6\mathcal{H}^2 = \hat{a}^2\rho
    \qquad\text{and}\qquad
    3(\mathcal{H}' + \mathcal{H}^2) = -\hat{a}^2p
\end{equation}
with, as usual 
\begin{equation}
\rho = 1/(2\hat{a}^2)(\phi')^2 + V
\qquad\text{and}\qquad
p = 1/(2\hat{a}^2)(\phi')^2 - V.
\end{equation}
These equations imply, in particular, some useful identities, such as the following
\begin{equation}\label{eq:pre_20}
    3(\mathcal{H}' - \mathcal{H}^2) + (\phi')^2 = 0.
\end{equation}
In the following, we still use the slow-roll parameters $\varepsilon$ and $\delta$ defined in \eqref{eq:pre_290}. But their potential counterparts $\varepsilon_V$ and $\eta_V$ have to be adapted, because the coefficients in the Friedmann equations are not the same in 5D, see \eqref{eq:pre_39}. In order to keep $\varepsilon\simeq \varepsilon_V$ and $\delta\simeq\eta_V-\varepsilon_V$ in the slow-roll limit, we take
\begin{equation}
    \varepsilon_V = \frac{3}{4}\biggl(\frac{V_{\phi}}{V}\biggr)^2 \quad\text{and}\quad \eta_V = \frac{3}{2}\frac{V_{\phi\phi}}{V}.
\end{equation}

\subsection{The scalar perturbations}

The general formalism for the study of cosmological perturbations in brane world theories has been studied extensively (see for instance \cite{vandeBruck:2000ju, Delfin:2023fcq} and reference therein). As we are interested here in inflation in a 5D quasi de Sitter space, our framework is just a direct generalization of the 4D case discussed in the previous section. Namely, we perturb the aforementioned background evolution just like in \eqref{eq:pre_7} 
\begin{eqnarray}\label{eq:pre_11}
ds^2 = \hat{a}^2(\tau)(&-&(1+2\Phi)d\tau^2 + 2B_idx^id\tau + 2Cdyd\tau  \nonumber\\
&+& ((1 - 2\Psi)\delta_{ij} + E_{ij})dx^idx^j + 2F_idx^idy + (R_0^2 - 2\Xi)dy^2)
\end{eqnarray}
where, again, we can further decompose $B_i, E_{ij}$ and $F_i$ into scalar, vector and tensor modes
\begin{equation}\label{eq:pre_26}
    B_i = \partial_i B + C_i
    ,\quad
    E_{ij} = 2\partial_{i}\partial_j E + 2\partial_{(i}E_{j)} + h_{ij}
    ,\quad
    F_i = \partial_i F + G_i.
\end{equation}
So we now have eight scalar perturbations $(\delta\phi, \Phi, \Psi, \Xi, B, C, E, F)$, three vector perturbations $(C_i, E_i, G_i)$ and one tensor perturbation $h_{ij}$. Under a 5D diffeomorphism with parameter $\xi^\mu = (\xi^t, \partial^i\xi_x + \tilde\xi^i, \xi^y)$, the perturbations transform following $\delta g_{\mu\nu}\rightarrow\delta g_{\mu\nu}-2\nabla_{(\mu}\xi_{\nu)}$
\begin{align}\label{eq:pre_10}
&\Phi\rightarrow\Phi-\mathcal{H}\xi^{t}-(\xi^{t})',\quad
\Psi\rightarrow\Psi+\mathcal{H}\xi^{t},\quad
\Xi\rightarrow \Xi+R_{0}^{2}\mathcal{H}\xi^{t}+R_{0}^2\partial_{y}\xi^{y},\nonumber\\
&B\rightarrow B+\xi^{t}-(\xi^x)',\qquad
C\rightarrow C+\partial_{y}\xi^{t}-R_{0}^2(\xi^{y})',\qquad
E\rightarrow E-\xi^x,\\
&F\rightarrow F-R_{0}^2\xi^{y}-\partial_{y}\xi^x,\quad
C^{i}\rightarrow C^{i}-(\tilde{\xi}^{i})',\quad
E^{i}\rightarrow E^{i}-\Tilde{\xi}^{i},\quad
G^{i}\rightarrow G^{i}-\partial_{y}\Tilde{\xi}^{i}.\nonumber
\end{align}
When computing these transformations, one has to take care of factors of $a(\tau)$ arising when lowering indices from $\xi^\mu$ to $\xi_\mu$. Besides the metric, the inflaton scalar field perturbation $\delta\phi$ transforms as before
\begin{equation}
\delta \phi\rightarrow\delta\phi-\phi'\xi^{t}.
\end{equation}
The gauge freedom allows to put three scalar perturbations to zero, using transformations with $(\xi^t, \xi^x, \xi^y)$. A possible choice is $\delta\phi = E = F = 0$ that we use in the following.

\subsubsection{Einstein equations}

We write here the 5D Einstein equations for the scalar perturbations of the metric \eqref{eq:pre_11} in the gauge $\delta\phi = E = F = 0$. The corresponding $00$ component is
\begin{eqnarray}\label{eq:pre_12}
R_0^2(3\mathcal{H}(3\Psi' + 4\mathcal{H}\Phi) - \Delta(2\Psi - 3 \mathcal{H} B))&&\nonumber\\
- \Delta \Xi-3 \partial_y^2\Psi+3\mathcal{H} \partial_y C+3 \mathcal{H} \Xi' &=& R_{0}^{2}(\phi')^{2}\Phi,
\end{eqnarray}
the $0i$ and $0y$ components are
\begin{eqnarray}\label{eq:pre_13}
R_0^2(4 \Psi' + 6\mathcal{H}\Phi)- \partial_y^2 B + \partial_y C +2 \Xi' &=& 0,\\\label{eq:pre_14}
\Delta(C-\partial_y B) - 6\partial_y\Psi' - 6\mathcal{H}\partial_y\Phi  &=& 0,
\end{eqnarray}
the $ii$ component gives
\begin{eqnarray}\label{eq:pre_15}
R_0^2 (2\Psi '' + 6\mathcal{H}\Psi' + 3\mathcal{H}\Phi' + 6 (\mathcal{H}'+\mathcal{H}^2) \Phi)&& \nonumber\\
+3 \mathcal{H}\partial_y C +3 \mathcal{H} \Xi'+(\partial_y C'+\Xi'')+\partial_y^2(\Phi -2\Psi)&&\nonumber\\
   +2/3\Delta(R_0^2(B' + 3 \mathcal{H} B + \Phi - \Psi) - \Xi) &=& -R_{0}^{2}(\phi')^{2}\Phi,
\end{eqnarray}
the $ij$ and $iy$ components give
\begin{eqnarray}\label{eq:pre_16}
R_0^2 (B' + 3 \mathcal{H} B + \Phi -\Psi ) - \Xi &=& 0,\\\label{eq:pre_17}
3 \mathcal{H} (\partial_y B + C)+\partial_y(B'+2 \Phi -4 \Psi ) + C' &=& 0,
\end{eqnarray}
and finally, the $yy$ component is
\begin{equation}\label{eq:pre_18}
6 (\mathcal{H}'+\mathcal{H}^2)\Phi +3 \mathcal{H} (\Delta B+\Phi '+3 \Psi ')+\Delta B'+\Delta \Phi - 2 \Delta \Psi +3 \Psi '' = -(\phi')^{2}\Phi.
\end{equation}
Here, $\Delta$ is the three dimensional Laplacian. In \eqref{eq:pre_12}, \eqref{eq:pre_13}, \eqref{eq:pre_15}, \eqref{eq:pre_16}, the terms with $R_0^2$ factorised are very similar to their four dimensional counterparts in \eqref{eq:pre_19} rewritten in conformal time, up to some numerical factors. Note also that the last line of \eqref{eq:pre_15}, with a $\Delta$ factorised, vanishes due to \eqref{eq:pre_16}.We now go to Fourier space, taking into account the compactness of the $y$ coordinate,
\begin{equation}\label{eq:pre_44}
A(\tau, x, y) = \int d^3k \sum_n A_{n}(k,\tau) e^{ikx}e^{in y},
\end{equation}
where $A$ stands for any 5D field. This amounts to replacing all $\Delta$ by $-k^2$ and the $\partial_y$ by $in$ in the previous equations, along with adding indices of $n$ everywhere, that we will omit in order to avoid cluttering the equations. Now, following the four dimensional case, three of the seven equations should be constraints on fields, two of them should boil down to equations on the background, and only two should remain genuine propagation equations.\\

Indeed, as a first step, we use \eqref{eq:pre_12}, \eqref{eq:pre_13} and \eqref{eq:pre_14} to express $(\Phi, B, C)$ as a function of the other perturbations. For this, note that \eqref{eq:pre_13} and \eqref{eq:pre_14} span a close, linear system on the variables $\Phi$ and $C - \partial_y B$, that can be solved to give
\begin{equation}\label{eq:pre_22}
    \Phi = -\frac{(3n^2 + 2R_0^2 k^2) \Psi'+ k^2\Xi'}{3\mathcal{H}(n^2 + R_0^2k^2)}
    \qquad\text{and}\qquad
    C - in B = -\frac{2in(R_0^2\Psi' - \Xi')}{n^2 + R_0^2k^2}.
\end{equation}
We can then use \eqref{eq:pre_12} to obtain $C$ and $B$ separately, but their expressions are not very enlightening.\\

The second step is to put the expressions for $(\Phi, B, C)$ back into the remaining equations \eqref{eq:pre_15}-\eqref{eq:pre_18}. We find that, as expected, two of these equations boil down to equations on the background which are therefore trivially satisfied. To be more precise, we find:
\begin{eqnarray}
    2i\times \eqref{eq:pre_15} + n \times \eqref{eq:pre_17} &=& 2iR_0^2(3 (\mathcal{H}'-\mathcal{H}^2)+ (\phi')^2))\Phi \\
    k^2\times \eqref{eq:pre_15} - n^2 \times \eqref{eq:pre_18} &=& -(n^2 - R_0^2k^2 )(3(\mathcal{H}'-\mathcal{H}^2)+ (\phi')^2))\Phi
\end{eqnarray}
so that both combinations vanish due to the background equation \eqref{eq:pre_20}. In other words, \eqref{eq:pre_17} and \eqref{eq:pre_18} are not independent, and we are left with two equations for $\Psi$ and $\Xi$. We now introduce a new variable $\Theta\equiv\Psi-\Xi/R_{0}^{2}$, for which we find
\begin{eqnarray}
    R_0^2\times\eqref{eq:pre_15} + n^2\times\eqref{eq:pre_16} = &-& R_0^4/\mathcal{H}(3(\mathcal{H}'-\mathcal{H}^2)+ (\phi')^2))\Psi'\nonumber\\
    &+& \text{terms that depend only on $\Theta$}.
\end{eqnarray}
Using \eqref{eq:pre_20} again, the first line vanishes and we get the following simple equation for $\Theta$
\begin{equation}\label{eq:Theta}
    \Theta'' + 3\mathcal{H}\Theta' + \biggl(k^2 + \frac{n^2}{R_0^2}\biggr)
    \Theta = 0,
\end{equation}
which is the equation of motion of a 5D massless minimally coupled scalar field around a time-dependent background, see \eqref{eq:pre_21}, reduced to one-dimensional equation for the time dependence of the Fourier coefficients of square mass $(k^2 + n^2/R_0^2)/\hat{a}^2$. Finally, from \eqref{eq:pre_15}, one can obtain a less simple equation on $\Psi$. Alternatively, introducing yet another variable $\Omega=((3n^2 + 2R_0^2 k^2) \Psi + k^2\Xi)/(3(n^2 + R_0^2 k^2))$ (note that $\Omega' = -\mathcal{H}\Phi$, with $\Phi$ as in \eqref{eq:pre_22}), we can combine the previous equations to obtain a simpler equation on $\Omega$ only
\begin{equation}\label{eq:pre_23}
    \Omega'' + \biggl(3\mathcal{H} + \frac{2(\mathcal{H}')^2 - \mathcal{H} \mathcal{H}''}{\mathcal{H}^3 - \mathcal{H} \mathcal{H}'}\biggr)\Omega' + \biggl(k^2 + \frac{n^2}{R_0^2}\biggr)
    \Omega = 0.
\end{equation}
Actually, this equation can be put in a more suggestive form, reminiscent of \eqref{eq:pre_24}
\begin{equation}\label{eq:pre_25}
    \frac{\hat{a}^{3}(\phi')^2}{\mathcal{H}^2}\Omega'' + \frac{d}{d\tau}\biggl(\frac{\hat{a}^{3}(\phi')^2}{\mathcal{H}^2}\biggr)\Omega' + \frac{\hat{a}^{3}(\phi')^2}{\mathcal{H}^2}\biggl(k^2 + \frac{n^2}{R_0^2}\biggr)\Omega = 0.
\end{equation}
So at the end, both \eqref{eq:Theta} and \eqref{eq:pre_25} can be identified as equations of motion following from
\begin{eqnarray}\label{eq:pre_28}
    \mathcal{S} = \frac{1}{2}\int d\tau d^3k \sum_n\biggl\{
    \hat{a}^{3}\biggl((\Theta')^2 - \biggl(k^2 + \frac{n^2}{R_0^2}\biggr)\Theta^2\biggr)&&\nonumber\\
    +\frac{\hat{a}^{3}(\phi')^2}{\mathcal{H}^2}\biggl((\Omega')^2 - \biggl(k^2 + \frac{n^2}{R_0^2}\biggr)\Omega^2\biggr)&\biggr\}&.
\end{eqnarray}
Note that the only difference between $\Theta$ and $\Omega$ is the factor of $(\phi')^{2}/\mathcal{H}^{2}=3\varepsilon$ in the $\Omega$ part, where $\varepsilon$ is the Hubble slow-roll parameter, $\varepsilon = -\dot H / H^2 = 1-\mathcal{H}'/\mathcal{H}^2$, see \eqref{eq:pre_20}.

\subsubsection{Power spectrum}

The action \eqref{eq:pre_28} has a form similar to \eqref{eq:pre_241} and can lead to equations similar to \eqref{eq:pre_3} upon defining 
\begin{equation}\label{eq:pre_40}
y\equiv \hat{a}^{3/2},\quad
z\equiv \hat{a}^{3/2}\phi'/\mathcal{H}, \quad 
\theta\equiv y\Theta, \quad
\omega \equiv z\Omega \quad\text{and}\quad
m_{k, n}^2 \equiv k^2 + n^2/R_0^2\,.
\end{equation}
We then obtain two copies of the Mukhanov-Chibisov equation as in \eqref{eq:pre_4}, see also \eqref{eq:pre_5}
\begin{equation}
\theta''_{k,n} + \biggl(m_{k,n}^2 - \frac{y''}{y}\biggr)\theta_{k,n} = 0 \quad\text{and}\quad
\omega''_{k,n} + \biggl(m_{k,n}^2 - \frac{z''}{z}\biggr)\omega_{k,n} = 0\,,
\end{equation}
where we restored the momentum dependence for clarity.
At linear order in the slow-roll parameters, recall from \eqref{eq:pre_290} to \eqref{eq:pre_29} that $\mathcal{H} \simeq -(1+\varepsilon)/\tau$ and $\hat{a}(\tau) \simeq -1/(H\tau^{1+\varepsilon})$, so we can obtain directly $y''/y$. We can also compute $z''/z$ as in \eqref{eq:pre_29}, and we get
\begin{equation}
    \frac{y''}{y}\simeq \frac{15/4 + 6\varepsilon}{\tau^2}
    \quad\text{and}\quad
    \frac{z''}{z}\simeq \frac{15/4 + 10\varepsilon - 4\delta}{\tau^2}.
\end{equation}
Thus, as in four dimensions, we end up with the following general solutions for $\theta_{k, n}$ and $\omega_{k, n}$, in terms of the Bessel functions $J_\nu$ and $Y_\nu$
\begin{eqnarray}\label{eq:pre_430}
    f_{k, n} = c_1^f \tau^{1/2}J_{\nu_f}(m_{k, n}\tau) + c_2^f \tau^{1/2}Y_{\nu_f}(m_{k, n}\tau),
    \quad\text{where $f = \theta$ or $\omega$},
\end{eqnarray}
and we have defined
\begin{equation}\label{eq:pre_43}
    \nu_\theta = 2+\frac{3\varepsilon}{2}
    \quad\text{and}\quad
    \nu_\omega = 2 + \frac{5\varepsilon}{2} - \delta,
\end{equation}
while the Bunch-Davies boundary conditions \eqref{eq:pre_30} give the same expressions for the constants $(c_1^f, c_2^f)$ in terms of $\nu_f$, as in \eqref{eq:pre_301}. We have thus obtained the solutions for $\theta_{k, n}$ and $\omega_{k, n}$, from which we can write the expression of the adiabatic curvature perturbation $\mathcal{R} = \Psi$
\begin{equation}\label{eq:pre_31}
    \mathcal{R}_{k, n} = \Omega_{k, n} + \frac{k^2 }{3 m_{k,n}^2}\Theta_{k, n} =
    \frac{1}{\sqrt{3\varepsilon}\hat{a}^{3/2}}\biggl(\omega_{k, n} + \sqrt{\frac{\varepsilon}{3}}\frac{k^2}{m_{k, n}^2}\theta_{k, n }\biggr).
\end{equation}

As in the four dimensional case, after obtaining the mode functions, we take the super horizon limit $k\tau \ll 1$ to compute the power spectrum \eqref{eq:pre_27}. This corresponds to perturbations with wavelengths that exit the Hubble radius, which is constant during inflation (in cosmic time). These modes then remain constant until they re-enter the horizon, which grows in subsequent eras after the end of inflation. \\

In our five dimensional case, the analogue limit is $m_{k, n}\tau \ll 1$. Again, this corresponds to five dimensional modes that exit the Hubble radius, which is constant during inflation, when all spatial dimensions expand homogeneously. After inflation, only the three spatial dimensions are still expanding, while the 5th (dark) dimension is assumed to be stabilised~\cite{Anchordoqui:2023etp}. In this super horizon limit perturbations behave with a power law
\begin{equation}\label{eq:pre_41}
    f_{k, n} \simeq (1+i)\sqrt{\frac{2\tau}{\pi}}\frac{1}{(m_{k, n}\tau)^{\nu_f}}.
\end{equation}
Let us now consider a mode of given 3D momentum $k$. In \eqref{eq:pre_11}, we denoted by $\hat a(\hat t) \sim e^{H\hat t}$ the scale factor of the 5D universe. Following \cite{Anchordoqui:2023etp}, we also denote by $a(t) \sim t^\alpha$. 
Here, $\hat t$ and $t$ are respectively the 5D and 4D cosmic times.\\

After inflation, the Dark Dimension is stabilised and the universe follows its standard four-dimensional evolution, with $H= 1/(2t)$ and physical momenta $q = k/a \sim k/t^{1/2}$, so that modes of physical scales that exited the Hubble radius during inflation will re-enter. Note that after inflation, we consider $H = \dot a(t) / a(t)$, without the hats.\\

During inflation, the universe is five-dimensional, so that physical wave-lengths that should be compared to the Hubble radius are $1/\hat q$, with $\hat q = \hat k/\hat a$ and $\hat k = m_{k, n}$ in our notation. The limit $m_{k, n}\tau \ll 1$ is equivalent to $\hat k/ \hat a \ll H$, corresponding to physical scales that exit the Hubble radius well before the end of inflation. Since $\hat k/\hat a \sim \hat k / e^{H\hat t}$ during inflation \cite{Anchordoqui:2023etp} while $H$ is constant, a large band of scales do exit the horizon. \\

\begin{figure}[ht!]\begin{center}
\includegraphics[scale=0.3]{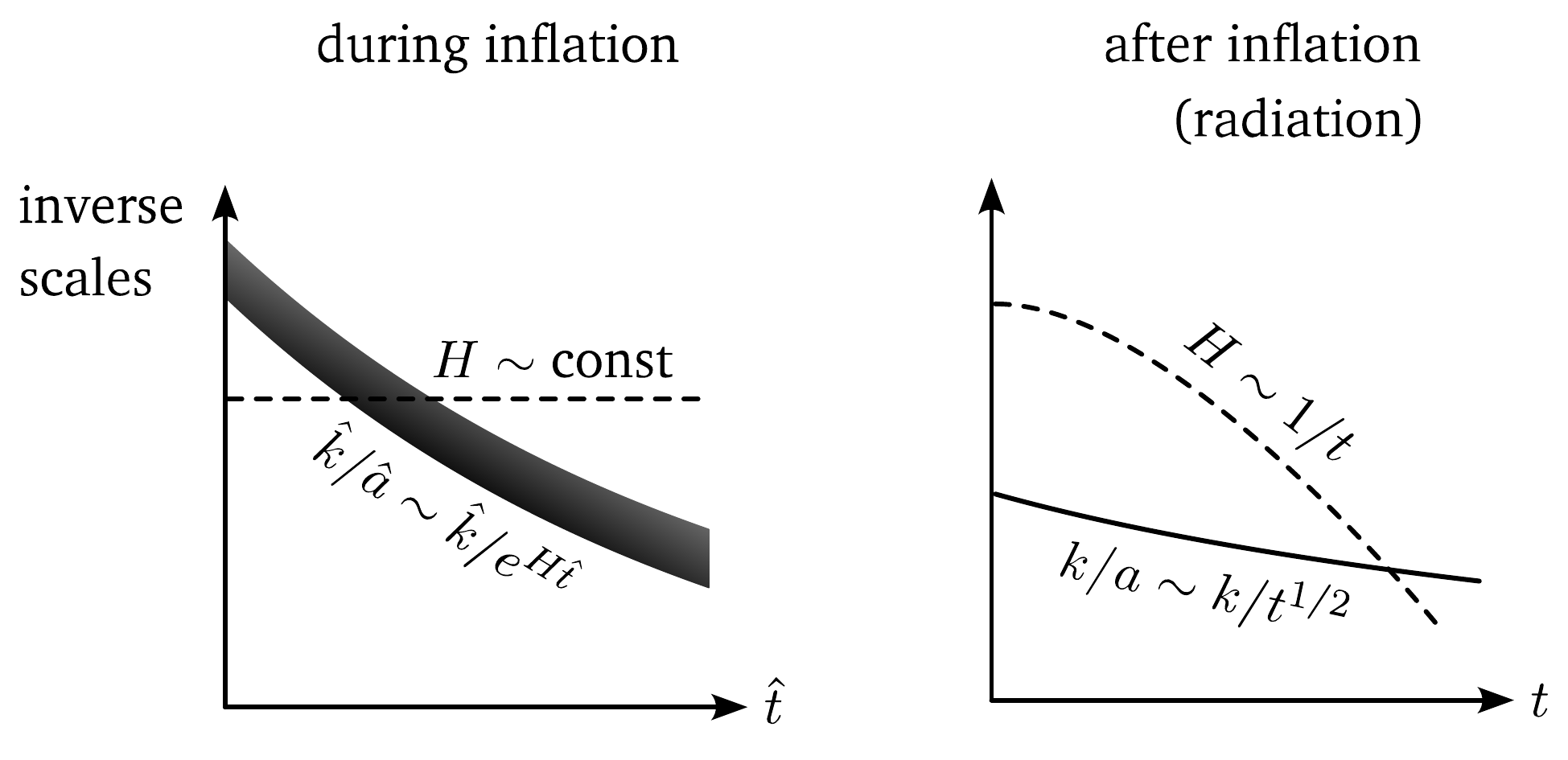}
\caption{Right panel shows evolution of physical scales and Hubble radius \emph{after} inflation. Since $H\sim 1/t$ while $q = k/a \sim k/t^{1/2}$, scales that exited the Hubble radius during inflation will eventually re-enter. Left panel shows evolution of physical scales and Hubble radius \emph{during} inflation. Since $H\sim\text{const}$ while $\hat q = \hat k /\hat a \sim \hat k/e^{H\hat t}$, a large band of scales exit the Hubble radius during inflation. A given 3D momentum $k$ corresponds to a tower of 4D momenta $\hat k$, within the black band. 
}
\label{fig:1}
\end{center}
\end{figure}

A given 3D momentum $k$ corresponds to a tower of 4D momenta $\hat k$, according to $k^2 = \hat k^2 - n^2/R_0^2$. So as $n$ increases with $k$ fixed, $\hat k$ increases. As a consequence, the very high modes of the tower fail to satisfy $\hat k/ \hat a \ll H$ and do not exit the horizon before the end of inflation. However, since 5D inflation last for about 40 e-folds there is a huge number of Kaluza-Klein (KK) modes that exit the horizon for a sufficiently large region of 3D momenta, which justifies the limit $m_{k, n}\tau \ll 1$ in \eqref{eq:pre_41}. Corrections from the expansion of the Bessel functions in \eqref{eq:pre_430} are expected to be irrelevant.
This is summarised in Figure \ref{fig:1}.

Note that we only observe perturbations in our four-dimensional universe which is located on the 3-brane at the origin of the 5th dimension. This corresponds to a summation over all KK modes. From a 4D perspective, a perturbation measured at a given spatial momentum $k$ corresponds to any momentum component along the extra dimension. Therefore, in order to evaluate the four-dimensional power spectrum of $\mathcal{R}$, we have to sum over all KK momenta
\begin{equation}\label{eq:pre_32}
    \mathcal{P}_\mathcal{R} = \frac{k^3}{2\pi^2}\sum_n|\mathcal{R}_{k, n}|^2,
\end{equation}
where using \eqref{eq:pre_31}
\begin{equation}\label{eq:pre_35}
    \frac{k^3}{2\pi^2}|\mathcal{R}_{k, n}|^2 \simeq \frac{2H^3k^3\tau^4}{3\pi^3\varepsilon}\biggl(
    \frac{1}{(m_{k, n}\tau)^{2\nu_\omega}}
    +\frac{\varepsilon}{3}\frac{k^{4}}{m_{k,n}^{4}}\frac{1}{(m_{k, n}\tau)^{2\nu_\theta}}\biggr)\,,
 \end{equation}
since the cross term that comes naively by squaring $|\mathcal{R}_{k, n}|^2$ is absent, because the perturbations $\omega_{k, n}$ and $\theta_{k, n}$ are Gaussian random variables at this level. Evaluating the sum \eqref{eq:pre_32} requires computing sums of the form
\begin{equation}\label{eq:pre_34}
    \sum_n \frac{1}{(R_0m_{k, n})^{2\alpha}} = \sum_n \frac{1}{((R_0k)^2 + n^2)^\alpha},
\end{equation}
where $\alpha$ is close to $4$. If $\alpha$ is an integer greater than $1$, it is rather direct to establish that
\begin{equation}\label{eq:pre_33}
S_\alpha(x) = \sum_n \frac{1}{(x + n^2)^\alpha}  = \frac{(-1)^{\alpha-1}\pi}{\Gamma(\alpha)}\biggl(\frac{\partial}{\partial x}\biggr)^{(\alpha-1)}\biggl(\frac{\coth(\pi\sqrt{x})}{\sqrt{x}}\biggr).
\end{equation}
For non integer $\alpha$, writing
\begin{equation}
    S_{\alpha}(x)=x^{\lfloor\alpha\rfloor-\alpha}\sum\limits_{n}\frac{1}{(x+n^{2})^{\lfloor\alpha\rfloor}}\left(\frac{x}{x+n^{2}}\right)^{\alpha-\lfloor\alpha\rfloor},
\end{equation}
where $\lfloor\alpha\rfloor$ denotes the integer part of $\alpha$, we can bound $S_{\alpha}(x)$ as
\begin{align}\label{sumbounds}
    x^{\lfloor\alpha\rfloor-\alpha+1}S_{\lfloor\alpha\rfloor+1}(x)<S_{\alpha}(x)\leqslant x^{\lfloor\alpha\rfloor-\alpha}S_{\lfloor\alpha\rfloor}(x).
\end{align}
In our case $\alpha-\lfloor\alpha\rfloor$ is proportional to the slow-roll parameters and is therefore very small. As a consequence, the upper bound given in \eqref{sumbounds} is a very good approximation; thus
\begin{equation}
    \sum_n \frac{1}{((R_0k)^2 + n^2)^\alpha} \simeq (R_0k)^{2(\lfloor\alpha\rfloor-\alpha)}\sum_n \frac{1}{((R_0k)^2 + n^2)^{\lfloor\alpha\rfloor}}.
\end{equation}
This formula works quite well for all $x > 0$. The error is proportional to $|\lfloor\alpha\rfloor-\alpha|$, and it is less than $1\%$ when $|\lfloor\alpha\rfloor-\alpha| < 0.06$ (recall that $|\lfloor\alpha\rfloor-\alpha|$ is linear in the slow-roll parameters). Using this formula in \eqref{eq:pre_32} yields
\begin{equation}\label{eq:pre_36}
    \mathcal{P}_\mathcal{R} \simeq \frac{2R_0H^3(R_0k)^3}{3\pi^3\varepsilon}\biggl(\biggl(\frac{k}{\hat{a}H}\biggr)^{4-2\nu_\omega}S_2((R_0k)^2)
    +\frac{\varepsilon}{3}\biggl(\frac{k}{\hat{a}H}\biggr)^{4-2\nu_\theta}(R_0k)^4S_4((R_0k)^2) 
    \biggr)
\end{equation}
where $S_2, S_4$ can be expressed using \eqref{eq:pre_33}, and in the limits $R_0 k \ll 1$ and $R_0k \gg 1$
\begin{eqnarray}\label{eq:pre_45}
    S_{\lfloor\alpha\rfloor}((R_0k)^2) \underset{R_0k \ll 1}{\simeq} \frac{1}{(R_0k)^{2\alpha}} ,\quad
    S_{\lfloor\alpha\rfloor}((R_0k)^2) \underset{R_0k \gg 1}{\simeq}\frac{\sqrt{\pi}\Gamma(\alpha-1/2)}{\Gamma(\alpha)} \frac{1}{(R_0k)^{2\alpha - 1}}.
\end{eqnarray}
As a result, the power spectrum in these two limits is
\begin{equation}\label{eq:pre_37}
    \mathcal{P}_\mathcal{R} \underset{R_0k \ll 1}{\simeq} \frac{2H^3}{3\pi^3\varepsilon k}\biggl(\biggl(\frac{k}{\hat{a}H}\biggr)^{2\delta - 5\varepsilon}+\frac{\varepsilon}{3}\biggl(\frac{k}{\hat{a}H}\biggr)^{-3\varepsilon}\biggr),
\end{equation}
and
\begin{equation}\label{eq:pre_47}
    \mathcal{P}_\mathcal{R} \underset{R_0k \gg 1}{\simeq} \frac{R_0H^3}{3\pi^2\varepsilon}\biggl(\biggl(\frac{k}{\hat{a}H}\biggr)^{2\delta - 5\varepsilon}+\frac{5\varepsilon}{24}\biggl(\frac{k}{\hat{a}H}\biggr)^{- 3\varepsilon}\biggr).
\end{equation}

For completeness, the expressions of $S_2((R_0k)^2)$ and $S_4((R_0k)^2)$ in \eqref{eq:pre_36} are
\begin{equation}
    S_2 = \frac{\pi}{2 x^{3/2}}  (\coth (u)+u\ \mathrm{csch}^2(u)),
\end{equation}
with $x = (R_0k)^2$ and $u = \pi\sqrt{x}$, and
\begin{equation}
    S_4 = \frac{\pi}{48 x^{7/2}}  (2 u^3 \mathrm{csch}^4 (u)+15 \coth (u)+u (4 u^2 \coth ^2(u)+12 u \coth (u)+15) \mathrm{csch}^2(u)).
\end{equation}

\subsubsection{Short discussion}

In the end, we recover the result of \cite{Anchordoqui:2023etp}, up to the normalisation factor $1/(3\varepsilon)$ which determines the amplitude, spectral tilt in the region $R_0k \gg 1$
\begin{equation}\label{scalartilt}
n_\mathcal{R}-1 = 2\delta - 5\varepsilon = 2\eta_V - 7\varepsilon_V.
\end{equation}
In the opposite limit $R_0k\ll 1$, the spectral tilt becomes $n_\mathcal{R}-1  = 2\eta_V - 7\varepsilon_V -1$, and the spectrum is no longer scale invariant.\\

The contribution of $\Omega$ to $\mathcal{R}$ in \eqref{eq:pre_31} comes with a factor of $1/\sqrt{\varepsilon}$ that becomes $1/\varepsilon$ in the power spectrum. By analogy with the four-dimensional case, see \eqref{eq:pre_27}, we can say that $\Omega$ is the contribution of the inflaton to the adiabatic curvature perturbation. Conversely, the contribution of $\Theta$ to $\mathcal{R}$ does not come with this factor, and gives the second term in \eqref{eq:pre_35}. In analogy with the tensor perturbations in \eqref{eq:pre_38}, one can say that this is a 5D gravitational contribution to the curvature perturbation that comes from the radion which is the second scalar, besides the inflaton, from a 4D perspective. However, in \eqref{eq:pre_37}, \eqref{eq:pre_47}, we can see that this contribution is proportional to $\varepsilon$ and thus subleading in the power spectrum.\\

At the end of the 5D inflation, the radion acquires a runaway potential of quintessence type and a stabilization mechanism of the fifth dimension is most likely required. This is discussed in \cite{Anchordoqui:2023etp}, where some possible contributions to the scalar potential are presented. In general, there are three additional contributions, besides the one arising from the 5D vacuum energy at the minimum of the inflaton potential, corresponding to 3-brane tensions, kinetic gradients of bulk fields and their Casimir energy. It is worth mentioning that the brane tension does not modify the bulk equations of motion, while bulk fields have vanishing expectation values during inflation, which occurs around a flat region of the inflaton potential away from its minimum. Finally, the Casimir energy falls off exponentially for bulk masses larger than the compactification scale and with a power much faster than the 5D cosmological constant when they are light, thus remaining negligible during inflation. 
Consequently, all these contributions, together with the 5D cosmological constant, can stabilize the fifth dimension without affecting the 5D inflation period discussed in our paper.

\section{Isocurvature perturbations}

When only one scalar degree of freedom is involved in inflation, perturbations are adiabatic. This implies that after inflation, the relative density of different matter components is constant; only the total density is perturbed. Intuitively, if there is only the inflaton, it decays in the same way everywhere to produce the thermal bath, up to density fluctuations. \\

When more scalar fields are involved, this is not anymore true because there can be perturbations leaving the total density unperturbed, called isocurvature or entropy perturbations ($\mathcal{S}$). If such a perturbation occurs between two fields, one of them decaying into a component that, for instance, does not thermalise with the rest of the bath (e.g. dark matter after decoupling), it leaves an imprint in the relative density of different components, such as matter $(m)$ and radiation $(\gamma)$
\begin{equation}
    \mathcal{S}_m = \frac{\delta\rho_m}{\rho_m} - \frac{3}{4}\frac{\delta\rho_\gamma}{\rho_\gamma} \neq 0.
\end{equation}
Such fluctuations would have an effect on the CMB. So far, it is subdominant. Isocurvature perturbations are characterised by their relative amplitude $\beta_\mathrm{iso}$ to the curvature perturbations
\begin{equation}\label{eq:pre_42}
    \beta_\mathrm{iso} = \frac{\mathcal{P}_\mathcal{S}}{\mathcal{P}_\mathcal{R}+\mathcal{P}_\mathcal{S}} < 0.038,
\end{equation}
constrained by Planck observations \cite{Planck:2018jri}.\\

As mentioned above, in our setup, from the 4D perspective there is an extra scalar besides the inflaton that drives 5D inflation: the so-called radion stemming from the 4D decomposition of the 5D metric. This may lead to
isocurvature perturbations that have to be addressed. Such perturbations are rather hard to compute precisely, as they depend on details about the reheating and how the different fields decay after inflation. To estimate their magnitude, we compute the total entropy perturbation \cite{Gordon:2000hv}
\begin{equation}
    \mathcal{S}=\mathcal{H}\left(\frac{\delta p}{p'}-\frac{\delta\rho}{\rho'}\right).
\end{equation}
Using $\delta p = \delta \rho = -(\phi')^2\Phi/\hat{a}^2$ and equations \eqref{eq:pre_39}, \eqref{eq:pre_20}, we get
\begin{equation}\label{eq:pre_46}
    \mathcal{S} = \biggl(\frac{-6\mathcal{H}^3 + 4\mathcal{H}'\mathcal{H}+\mathcal{H}''}{4(2\mathcal{H}^{3}-\mathcal{H}'')}\biggr)\Phi.
\end{equation}
Recalling that $\varepsilon = 1 - \mathcal{H}'/\mathcal{H}^2$, we can also evaluate $\varepsilon' = 2\mathcal{H}\varepsilon(\varepsilon-\delta)$, and
\begin{equation}
    \mathcal{H}^2 - \mathcal{H}' \simeq \mathcal{H}^2\varepsilon
    \quad\text{and}\quad
    2\mathcal{H}^3 - \mathcal{H}'' \simeq 4\mathcal{H}^3\varepsilon - 2\mathcal{H}^3\varepsilon\delta.
\end{equation}
Thus, at leading slow-roll order and using the definitions \eqref{eq:pre_40} we get
\begin{equation}
    \mathcal{S} \simeq -\frac{\Phi}{2} \simeq \frac{\Omega'}{2\mathcal{H}} \simeq \frac{\Omega}{2\mathcal{H}}\biggl(\frac{\omega'}{\omega} - \frac{z'}{z}\biggr)\,.
\end{equation}
Using now the expression of $\omega$ in the super horizon limit \eqref{eq:pre_41}, we can see that $\omega'/\omega\simeq(1/2-\nu_\omega)/\tau$, and $z'/z \simeq \mathcal{H}(3/2 + \varepsilon - \delta)$, leading to
\begin{equation}
    \mathcal{S} \simeq \frac{3\varepsilon}{4}\Omega \quad\rightarrow\quad \mathcal{P}_\mathcal{S} \simeq \frac{9\varepsilon^2}{16}\mathcal{P}_\mathcal{R} \quad\rightarrow\quad \beta_{\mathrm{iso}} \simeq \frac{9\varepsilon^2}{16}.
\end{equation}
Thus, we obtain that $\beta_\mathrm{iso}$ is second order in the slow-roll parameter $\varepsilon$, which easily evades the constraint \eqref{eq:pre_42}. But again, the total entropy perturbation is at best a rough estimate of the isocurvature perturbations. A more thorough analysis would be necessary to compute them more precisely, but this goes beyond the scope of this paper.\\

Note that $\mathcal{S}$ being proportional to $\Phi$ in \eqref{eq:pre_46} is a consequence of our gauge choice $\delta\phi = 0$. In ``genuine" four-dimensional multifield inflation, it is not possible to set both $\delta\phi_1 = 0$ and $\delta\phi_2 = 0$ with a gauge choice. This leads to an additional contribution in \eqref{eq:pre_46}, from the entropy field, defined as $\delta s$ in \cite{Gordon:2000hv}, which is a linear combination of $\delta\phi_1$ and $\delta\phi_2$. In our case, this contribution is absent, and the total entropy perturbation ends up being suppressed.

\section{Tensor and vector perturbations}

\subsection{Tensor perturbations}

For the tensor perturbation $h_{ij}$ of the metric \eqref{eq:pre_11}, since by definition $h^i_i = 0$ and $\partial^ih_{ij} = 0$, the only non trivial components of the Einstein equation are the $ij$ ones, which read
\begin{equation}\label{tensoreq}
    h_{ij}''+3\mathcal{H}h_{ij}'+\biggl(k^{2}+\frac{n^{2}}{R_{0}^{2}}\biggr)h_{ij}=0.
\end{equation}
Therefore the two independent helicities of $h_{ij}$ satisfy \eqref{eq:Theta}, as expected, and can be studied in exactly the same way as $\Theta$. At the end, the power spectrum we obtain is as \eqref{eq:pre_37}, up to a factor of $2\times 4\times 3\varepsilon$ (the factor of $2$ comes from the two polarisations, the factor of $4$ comes from the normalisation of $h$, see \eqref{eq:pre_500} and discussion after \eqref{eq:pre_38}, and the factor of $3\varepsilon$ comes from the factor of $\phi'/\mathcal{H}$ in \eqref{eq:pre_28}), and the spectral tilt that becomes $4-2\nu_\theta$, see \eqref{eq:pre_43}
\begin{eqnarray}\label{eq:pre_48}
    \mathcal{P}_\mathcal{T} \underset{R_0k \ll 1}{\simeq} \frac{16H^3}{\pi^3 k}\biggl(\frac{k}{\hat{a}H}\biggr)^{- 3\varepsilon}
    \quad\text{and}\quad
    \mathcal{P}_\mathcal{T} \underset{R_0k \gg 1}{\simeq} \frac{8R_0H^3}{\pi^2}\biggl(\frac{k}{\hat{a}H}\biggr)^{- 3\varepsilon}.
\end{eqnarray}
It follows that the tensor-to-scalar ratio $r$ and the tensor spectral tilt $n_\mathcal{T}$ are
\begin{equation}
    r = 24\varepsilon = 24\varepsilon_V,
\end{equation}
and
\begin{equation}
    n_\mathcal{T} = -3\varepsilon_V \quad\text{for}\quad R_0k \gg 1, \qquad
    n_\mathcal{T} = -3\varepsilon_V -1 \quad\text{for}\quad R_0k \ll 1.
\end{equation}
The experimental constraint $r < 0.06$ reported by \cite{Planck:2018jri} then sets a limit $\varepsilon_V < 0.003$.
On the other hand, using the experimental value of the scalar spectral index $n_\mathcal{R}-1\simeq -0.04$ and the result \eqref{scalartilt}, one finds that $\eta_V$ should be in the range $[-0.02, -0.01]$.

\subsection{Vector perturbations}

In ordinary 4D inflation, vector perturbations are not sourced at linear order and are thus irrelevant. However, a five dimensional metric contains helicities $\pm1$ from a 4D point of view, which correspond to vector fields; these are part of the massive KK graviton excitations which couple to (and thus are sourced by) the brane energy-momentum tensor. 
For the vector perturbations $(C_i, E_i, G_i)$, we can choose a gauge where $E_i = 0$. Then the only non-trivial parts of the Einstein equations are the $0i$, $ij$ and $iy$ ones which give
\begin{equation}
R_{0}^{2}\Delta C_{i}+\partial_{y}^2C_{i}-\partial_{y}G_{i}' = 0, \qquad
R_0^2(3\mathcal{H} C_i +  C_i') - \partial_y G_i = 0,
\end{equation}
and
\begin{equation}
3\mathcal{H}(G_i' - \partial_yC_i) - \Delta G_i - \partial_y C_i' + G_i'' = 0.
\end{equation}
In Fourier space \eqref{eq:pre_44}, we can use the first equation to obtain
\begin{equation}
    C_i = -\frac{inG_i'}{n^2 + R_0^2k^2}.
\end{equation}
Then, the two other equations turn out to be both equivalent to 
\begin{equation}
    G_i'' + 3\mathcal{H}G_i' + \biggl(k^{2}+\frac{n^{2}}{R_{0}^{2}}\biggr)G_i = 0\,,
\end{equation}
which is the same as \eqref{tensoreq} and \eqref{eq:Theta}.\\

Thus, the power spectrum of primordial vector perturbations is the same as for the tensors, and twice the contribution of the radion $\Theta$ to the scalar perturbations, as expected from the fact that all of them emerge from 5D gravitons. Note however that in the orbifold case where the 5th dimension forms an interval $S^1/\mathbb{Z}_2$, which is standard in brane-world constructions, the vector zero mode as well as half of the massive modes are projected out by the $\mathbb{Z}_2$ action. Projecting out the zero-mode boils down to extract $1/(R_0k)^{2\alpha}$ from the sum in \eqref{eq:pre_34}. This affects only the result in the limit $R_0k \ll 1$ in \eqref{eq:pre_45}, that becomes
\begin{equation}
    S_\alpha ((R_0k)^2) - \frac{1}{(R_0k)^{2\alpha}} \underset{R_0k \ll 1}{\simeq} \zeta(2\alpha).
\end{equation}
Therefore
\begin{eqnarray}\label{eq:vectorpower}
    \mathcal{P}_\mathcal{V} \underset{R_0k \ll 1}{\simeq} \frac{4\pi R_0H^3}{45}(R_0 k)^3\biggl(\frac{k}{\hat{a}H}\biggr)^{- 3\varepsilon}
    \quad\text{and}\quad
    \mathcal{P}_\mathcal{V} \underset{R_0k \gg 1}{\simeq} \frac{4R_0H^3}{\pi^2}\biggl(\frac{k}{\hat{a}H}\biggr)^{- 3\varepsilon}.
\end{eqnarray}
So the power spectrum of vector perturbations has a reduced amplitude at large distances, in contrast with all other perturbations which are more important due to an approximate $1/k$ behaviour. On the other hand, vector perturbations could also lead to $B$-mode polarization of the CMB which should be added to $\mathcal{P}_{\mathcal T}$ found previously in \eqref{eq:pre_48}.
Note that the $\mathbb{Z}_2$ action also truncates the sum over the KK modes \eqref{eq:pre_32} to $n > 0$, effectively adding a factor of $1/2$ everywhere this sum appears. This factor was not included in the previous power spectra \eqref{eq:pre_37}, \eqref{eq:pre_47}, \eqref{eq:pre_48}, but is taken into account in \eqref{eq:vectorpower}.

\section{Conclusions}

Following the recent proposal \cite{Anchordoqui:2023etp}, that five-dimensional inflation can be used to generate a large extra-dimension in the micron range, compatible with the Dark Dimension proposal \cite{Montero:2022prj}, we have studied cosmological perturbations in this setup. The main difference with the standard four-dimensional inflation is a change of behaviour in the power spectra at large distance scales, corresponding to angles larger than about $10$ degrees. This corresponds to wave-lengths that were larger than the extra dimension during inflation. Observations of CMB so far do not distinguish whether this change of behaviour happens or not in the Universe. The expressions of the spectral tilts and scalar-to-tensor ratio as a function of the 5D inflaton potential slow-roll parameters are also slightly modified with respect to the 4D case. The radion induced contribution to the scalar power spectrum is suppressed by the first slow-roll parameter $\varepsilon$, as well as the vector, isocurvature, and entropy perturbations. The presence of this second scalar may also induce non-gaussianities less suppressed than in single field 4D inflation models which would be interesting to compute.

\newpage
\bibliography{bibliographie}

\end{document}